\documentclass{pasj00}
\usepackage[OT2,T1]{fontenc}

\def\inpress{in prss}

\def\arxiv#1{ (arXiv astro-ph/#1)}

\DeclareAbbreviation\AAHam{Astron. Abh. Hamburg. Sternw.}
\DeclareAbbreviation\AARv{Astron. Astrophys. Rev.}
\DeclareAbbreviation\AAS{American Astron. Soc. Meeting Abstracts}
\DeclareAbbreviation\AcA{Acta Astron.}
\DeclareAbbreviation\actaa{Acta Astron.}
\DeclareAbbreviation\Afz{Astrofizika}
\DeclareAbbreviation\AGAb{Astronomische Gesellschaft Abstract Ser.}
\DeclareAbbreviation\an{Astron. Nachr.}
\DeclareAbbreviation\AnAp{Annales d'Astrophysique}
\DeclareAbbreviation\AnTok{Tokyo Astron. Obs. Annals, Sec. Ser.}
\DeclareAbbreviation\Ap{Astrophysics}
\DeclareAbbreviation\ARep{Astron. Rep.}
\DeclareAbbreviation\ATel{Astron. Telegram}
\DeclareAbbreviation\ATsir{Astron. Tsirk.}
\DeclareAbbreviation\AcApS{Acta Astrophys. Sinica}
\DeclareAbbreviation\AstL{Astron. Lett.}
\DeclareAbbreviation\BaltA{Baltic Astron.}
\DeclareAbbreviation\BANS{Bull. of the Astron. Institutes of the Netherlands Suppl. Ser.}
\DeclareAbbreviation\BASI{Bull. Astron. Soc. India}
\DeclareAbbreviation\BeSN{Be Newslett.}
\DeclareAbbreviation\BHarO{Harvard Coll. Obs. Bull.}
\DeclareAbbreviation\CBET{Cent. Bur. Electron. Telegrams}
\DeclareAbbreviation\ChJAA{Chinese J. of Astron. and Astrophys.}
\DeclareAbbreviation\caa{Chinese J. of Astron. and Astrophys.}
\DeclareAbbreviation\CoAsi{Asiago Contr.}
\DeclareAbbreviation\CoSka{Contributions of the Astronomical Observatory Skalnat\'e Pleso}
\DeclareAbbreviation\GCN{GRB Coord. Netw. Circ.}
\DeclareAbbreviation\ErgAN{Erg. Astron. Nachr.}
\DeclareAbbreviation\ibvs{IBVS}
\DeclareAbbreviation\IEEEP{IEEE Proc.}
\DeclareAbbreviation\JAD{J. Astron. Data}
\DeclareAbbreviation\JAVSO{J. American Assoc. Variable Star Obs.}
\DeclareAbbreviation\JBAA{J. Br. Astron. Assoc.}
\DeclareAbbreviation\JPhCS{J. of Physics Conference Series}
\DeclareAbbreviation\JPSJ{J. Phys. Soc. Japan}
\DeclareAbbreviation\JSARA{J. of the Southeastern Assoc. for Research in Astron.}
\DeclareAbbreviation\LowOB{Lowell Obs. Bull.}
\DeclareAbbreviation\MitAG{Mitteil. der Astronom. Gesell. Hamburg}
\DeclareAbbreviation\MitVS{Mitteil. Ver\"{a}nderl. Sterne}
\DeclareAbbreviation\MmSAI{Mem. Soc. Astron. Ital.}
\DeclareAbbreviation\Msngr{Messenger}
\DeclareAbbreviation\NewA{New Astron.}
\DeclareAbbreviation\na{New Astron.}
\DeclareAbbreviation\NewAR{New Astron. Rev.}
\DeclareAbbreviation\NInfo{Nauchnye Informatsii}
\DeclareAbbreviation\OAP{Odessa Astron. Publ.}
\DeclareAbbreviation\Obs{Observatory}
\DeclareAbbreviation\OEJV{Open Eur. J. on Variable Stars}
\DeclareAbbreviation\PASA{Publ. Astron. Soc. Australia}
\DeclareAbbreviation\PASAu{Publ. Astron. Soc. Australia}
\DeclareAbbreviation\PAZh{Pis'ma AZh}
\DeclareAbbreviation\POBeo{Publ. de l'Observatoire Astronomique de Beograd}
\DeclareAbbreviation\PCCP{Phys. Chem. Chem. Phys.}
\DeclareAbbreviation\PhR{Phys. Rep.}
\DeclareAbbreviation\PVSS{Publ. Variable Stars Sect. R. Astron. Soc. New Zealand}
\DeclareAbbreviation\PZ{Perem. Zvezdy}
\DeclareAbbreviation\PZP{Perem. Zvezdy, Prilozh.}
\DeclareAbbreviation\QJRAS{QJRAS}
\DeclareAbbreviation\RA{Ricerche Astronomiche}
\DeclareAbbreviation\RMxAA{Rev. Mexicana Astron. Astrof.}
\DeclareAbbreviation\RvMA{Reviews of Modern Astron.}
\DeclareAbbreviation\SASS{Society for Astronom. Sciences Ann. Symp.}
\DeclareAbbreviation\Sci{Science}
\DeclareAbbreviation\SPIE{SPIE Proc.}
\DeclareAbbreviation\SvA{Soviet Astronomy}
\DeclareAbbreviation\SvAL{Soviet Astronomy Letters}
\DeclareAbbreviation\VeSon{Ver\"{o}ff. Sternw. Sonneberg}
\DeclareAbbreviation\VSOLJBul{VSOLJ Variable Star Bull.}
\DeclareAbbreviation\yCat{VizieR Online Data Catalog}
\DeclareAbbreviation\ZA{Z. Astrophys.}

\def\ASPConf#1#2{ASP Conf. Ser. #1, #2}

\def\PublisherCambridge{Cambridge: Cambridge University Press}

\def\PublisherASP{San Francisco: ASP}

\def\PublisherSpringer{Berlin: Springer-Verlag}

\newcounter{author}
\def\authorcount#1#2{\refstepcounter{author}\label{#1}
           \altaffiltext{\ref{#1}}{#2}}
\begin{document}
\SetRunningHead{P. Zemko, T. Kato}{Detection of Change in Supercycles in ER UMa}

\Received{200X/XX/XX}
\Accepted{200X/XX/XX}

\title{Detection of Change in Supercycles in ER Ursae Majoris}

\author{Polina~\textsc{Zemko},\altaffilmark{\ref{affil:Shugarov},}\altaffilmark{\ref{affil:Kyoto}}
    Taichi~\textsc{Kato},\altaffilmark{\ref{affil:Kyoto}}
    Sergei~Yu.~\textsc{Shugarov},\altaffilmark{\ref{affil:Shugarov},}\altaffilmark{\ref{affil:Shugarov2}}
    }

\authorcount{affil:Shugarov}{
   Sternberg Astronomical Institute, Moscow University, Universitetsky
   Ave., 13, Moscow 119992, Russia}
\email{polina.zemko@gmail.com}

\authorcount{affil:Kyoto}{
   Department of Astronomy, Kyoto University, Kyoto 606-8502}
\email{tkato@kusastro.kyoto-u.ac.jp}

\authorcount{affil:Shugarov2}{
   Astronomical Institute of the Slovak Academy of Sciences, 05960,
   Tatranska Lomnica, the Slovak Republic}

\KeyWords{accretion, accretion disks
     --- stars: dwarf novae
     --- stars: individual (ER UMa)
     --- stars: novae, cataclysmic variables}

\maketitle

\begin{abstract}
We examined data on about 20 years of observations of ER UMa available in AAVSO,
 VSNET, AFOEV, NSVS, VSOLJ databases together with published light curves. 
The obtained $O-C$ diagram revealed a systematic change of the supercycle
 (time interval between two successive superotbursts) within 43.6 and 59.2 d.
 The time-scale of this change variation is from 300 to $\sim$ 1900 d. The
 number of normal outbursts within the supercycles also varied from 4 to 6
 although no strong correlation between this number and supercycle length was
 found. We suggest that appearance of negative superhumps is responsible for
 the observed variations of number of normal outbursts. Our results
 generally confirms the expectations by the thermal-tidal instability theory. 
\end{abstract}

\section{Introduction}

  Dwarf novae (DNe) are a class of cataclysmic variables (CVs) ---
close binary systems consisting of a white dwarf as a primary component and a 
red or brown-dwarf secondary transferring matter via the Roche-lobe overflow.
 Because of the instabilities that take place in the accretion disks normal
 outbursts and less frequent superoutbursts with larger amplitudes take place
(see \cite{war95book}, \cite{hel01book}). DNe that show both normal 
outbursts and superutbursts are called the SU UMa-type stars. These stars
 have one distinguishing property --- light-curve modulation with the
 periods close to, but different from, the orbital one that appear near the 
superoutburst maxima and sometimes persist at other times
(\cite{vog80suumastars}). These modulations are called positive superhumps if
 their period is longer then the orbital one and negative if shorter. Positive
  superhumps are believed to arise from the periodic viscous dissipation
in a disk driven to resonant oscillation by the tidal field of the
secondary star, and precessing slowly in the prograde direction 
 (see \cite{whi88tidal}, 
\cite{hir90SHexcess}, \cite{lub91SHa}, \cite{smi07SH}, \cite{woo11v344lyr}) and
 negative superhumps arise from the precession of tilted accretion disk, 
respectively  (\cite{woo09negativeSH}, \cite{mon10disktilt}). 

SU UMa stars have two extreme subgroups with very long (about decades) and short
 (tens of days) supercycles. They are WZ Sge and ER UMa subtypes, respectively.
There are five ER UMa-type stars known by now: RZ LMi (\cite{nog95rzlmi}),
 DI UMa (\cite{kat96diuma}), ER UMa (\cite{kat95eruma}), V1159 Ori
 (\cite{rob95eruma}), IX Dra (\cite{ish01ixdra}) and all of them show extremely 
short supercycle (18.9 -- 54 d) and positive superhumps 
(\cite{ole08curverzlmi}). Moreover, ER UMa-type stars were known to show a
 pronounced stability of the supercycle length and outburst pattern, which can
 be particularly seen in the folded light-curves and $O-C$ diagrams presented
 in \citep{rob95eruma}. This stability and short supercycles can be explained
 as a result of a very high and constant mass-transfer rate in the framework of
 thermal tidal instability (TTI) model (\cite{osa95eruma}).

 However, long-term monitoring of V1159 Ori revealed secular changes in
 the supercycle and a hint of the periodic behavior of such variations 
(\cite{kat01v1159ori}). The TTI model suggests that this phenomenon is the
 consequence of a change in mass-transfer rate in the binary system, although
 the origin of these changes is still an open question. The hibernation
 scenario, or the theory of cyclic evolution, suggested by \citet{liv92hib},
 also assumes 
that mass-transfer rate in CVs can vary secularly and proposes irradiation of
 the secondary by the white dwarf which is still hot after the nova eruption
 as one of the possible mechanisms of an increased mass-transfer. However, the 
time-scale of $\dot{M}$ variation in this scenario is very long comparing to 
these observations. On the other hand for V1159 Ori, \citet{kat01v1159ori}
 proposed that solar-type cycles of the
 secondary which are possibly observed in cataclysmic variables
 (\cite{bia88CVcycle}, \cite{ak01CVcycle}) may be responsible for the
 phenomenon.

 Here we report that another one ER UMa-type star --- ER UMa itself, also shows 
variations of the supercycle. A discussion on the origin of this phenomenon is 
even more intriguing in the context of recent discovery by \citet{ohs12eruma}
 of persistent negative superhumps in this dwarf nova. The presence of negative 
superhumps was not taken into account in past studies of SU UMa-type stars,
 while this factor may strongly affect the outburst behavior. 
\citet{osa121504cyg} have also found that the presence of negative
 superhumps is
 correlated with normal outbursts, thus it is reasonable to investigate the
 possibility whether the presence of negative superhumps is related to the
 long-term variations of the outburst behavior.

A comprehensive study of long-term variations in ER UMa-type stars is therefore
 essential for better understanding the evolution of dwarf novae in general,
 in particular changes in mass transfer-rate, the mechanisms that lead to the
 appearance of negative superhumps and their effect on the outburst behavior.

\section{Observation and Data Analysis}

\subsection{Light-curves}
\label{subsec:1lc}

\begin{figure*}
\begin{center}
\FigureFile(150mm,150mm){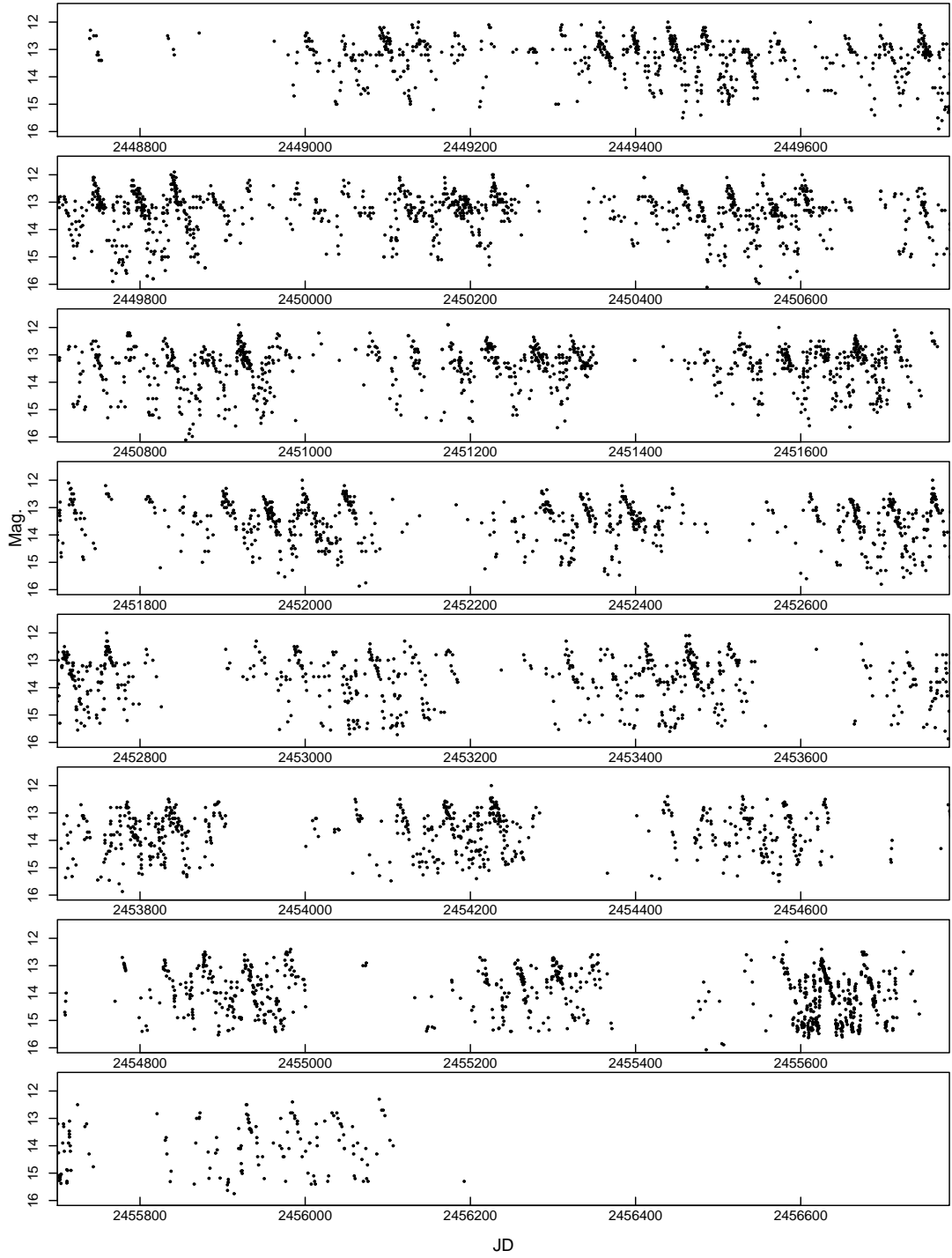}
\end{center}
\caption{Long-term light-curve of ER UMa from AAVSO, VSNET, AFOEV, NSVS and
 VSOLJ observations.}
\label{fig:tot}
\end{figure*}

We analyzed all the data on ER UMa available in American Association of
 Variable Stars Observers (AAVSO)\footnote{%
http://www.aavso.org/}, Variable Stars NETwork (VSNET)\footnote{%
http://www.kusastro.kyoto-u.ac.jp/vsnet/}, Association Fran\c{c}aise des 
Observateurs d'Etoiles Variables
 (AFOEV)\footnote{%
http://cdsarc.u-strasbg.fr/afoev/},
 Northern Sky Variability Survey (NSVS)\footnote{%
http://skydot.lanl.gov/nsvs/nsvs.php}
 and Variable Star Observers League in Japan (VSOLJ)\footnote{%
http://vsolj.cetus-net.org/}
 databases which represent more then 20 years of observations. The
earliest superoutburst was on 1992 April 27 (JD2448740) and
the latest one on 2012 June 12 (JD2456090.7). We also used observations
obtained by one of the authors (S. Shugarov) in the Star\'{a} Lesn\'{a}
 Observatory of Slovak Academy of Sciences, and by the group of E. Pavlenko in
 the Crimean Astrophysical Observatory.
A long-term light-curve of ER UMa from these observations 
is presented in figure \ref{fig:tot}.

\subsection{Moments of the superoutbursts and $O-C$ diagram}
\label{subsec:1oc}

In order to follow the supercycle change in ER UMa we plotted $O-C$ diagram
using the data mentioned. The moments of the superoutbursts and normal outbursts
were determined visually from the light curves. The mean period essential for 
plotting such diagram was determined by summing up the cycle counts and was 
chosen to be 48.7 d by the least-squares method. There were several gaps in
 observations usually 3-4 cycles long, particularly around the solar
 conjunction, which can be also seen in figure \ref{fig:tot}. The time
 intervals between two subsequent
detected superoutbursts in some such cases were large enough to bring 
considerable uncertainties in calculations of the number of cycles in such 
gaps. 
For every uncertain moment of the superoutburst we checked the two closest
 possible values of the cycle number, for each of them calculated the mean
 period for the few closest supercycles before, across and after the gap and 
regarded the variant as less probable if it leaded to the significant 
period jump. The list of the moments of superoutbursts, corresponding 
supercycle count --- $E$ (the number of superoutbursts passed since JD2448740), 
supercycle length, measured by the linear fitting of successive five epochs of
 $O-C$ diagram and the number of normal outbursts observed within the 
supercycle is presented in the table \ref{tab:data}.

\subsection{Normal outbursts phases}
\label{subsec:1no}

\begin{figure}
\includegraphics[width=85mm,height=80mm]{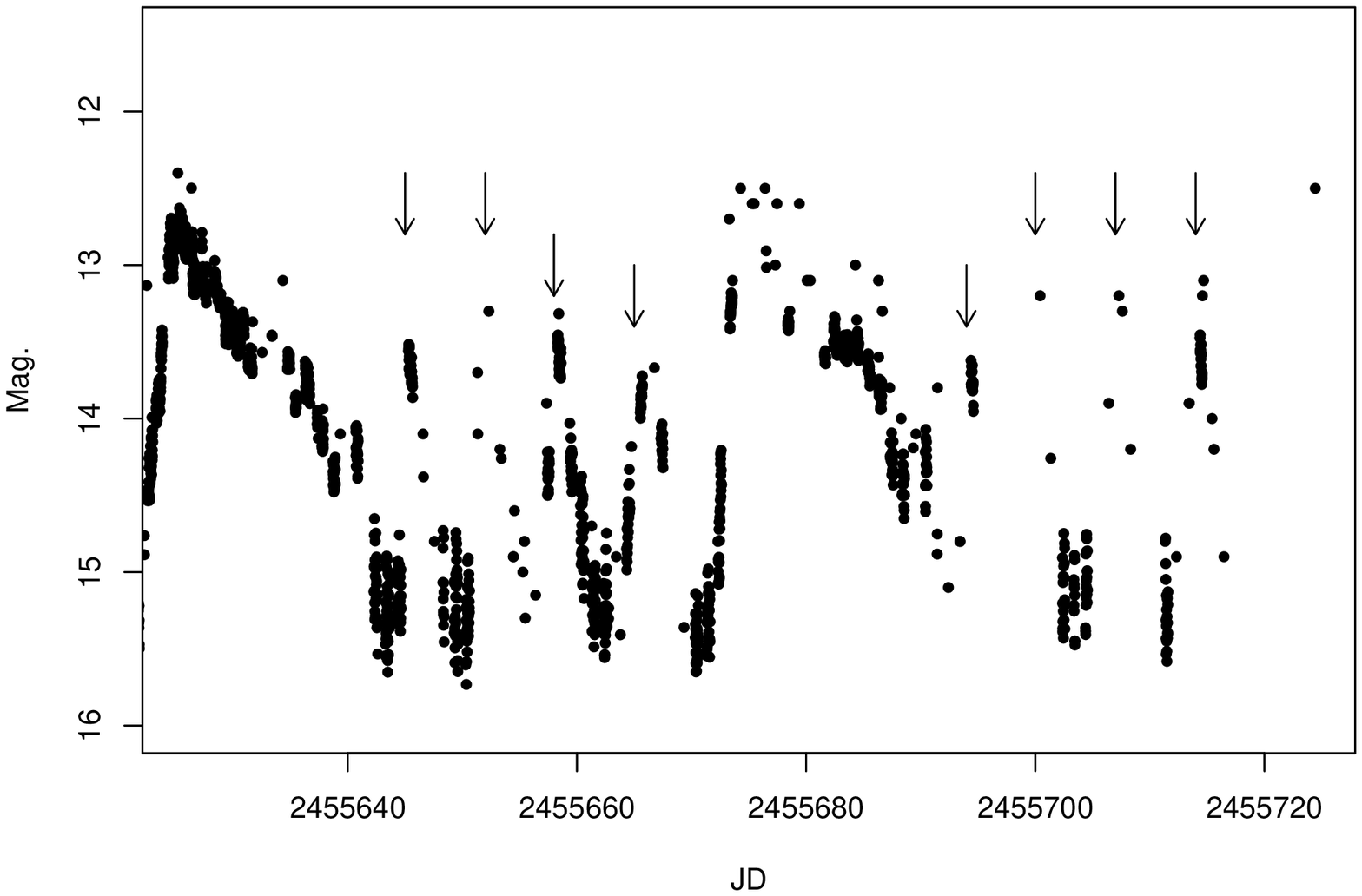}
\hfill
\includegraphics[width=85mm,height=80mm]{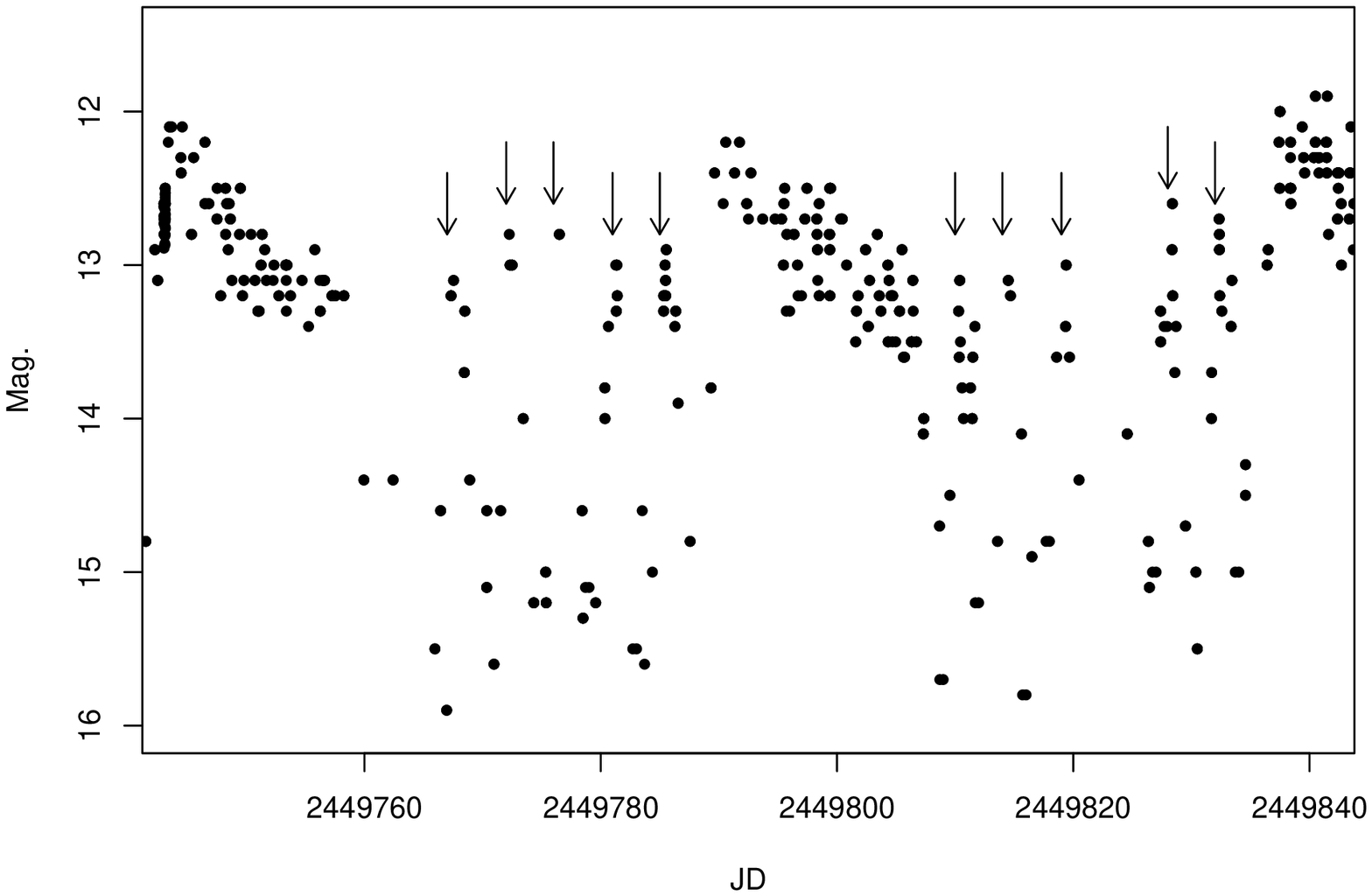}
\hfill
\parbox{80mm}{\caption{Example of the L type normal outburst behavior of ER UMa
 (March - June 2011)}\label{fig:l}}
\parbox{80mm}{\caption{Example of the S type normal outburst behavior of ER UMa
(January - May 1995)}\label{fig:s}}
\end{figure}

We noticed that ER UMa shows two types of outburst behavior with 3-4 normal 
outbursts within the supercycle observed and with 5 or more. Examples of these
 types can be seen in figure \ref{fig:l} and figure \ref{fig:s}. Following 
\citet{osa121504cyg} and \citet{sma85vwhyi} here we use Type S for
 the stages with many normal outbursts and short intervals and Type L with
 reduced numbers of normal outbursts and long intervals.
 Because of the gaps in observations and because the data were not very precise
 for the most part of supercycles it was possible to determine only approximate
numbers of normal outbursts within them, or even only lower limits. 
In the table \ref{tab:data}
 we give the probable numbers, as ``3-4'' for the cases were it was
 impossible to choose the unique value,
 and lower limits, as ``$>$4'', for the supercycles, where for example four
 normal outbursts were detected but the significant part of the supercycle
 remained unobserved, and we could not say whether there were other normal
 outbursts or
 not.

\begin{longtable}{cll|c|c|cll|c|c}
\caption{Moments of the superoutbursts, supercycle lengths ($T_{\rm{s}}$) and 
the number of normal outbursts within the supercycle ($N_{\rm{NO}}$). Since the
 moments of 
superoutbursts were determined from
several databases, multiple entries for the same $E$ sometimes occur.}\label{tab:data}
\hline
$E$ & Date & JD & $T_{\rm{s}}$ (d) & $N_{\rm{NO}}$ & E & Date & JD & $T_{\rm{s}}$ (d)& $N_{\rm{NO}}$\\
\hline
\hline
\endfirsthead
\hline
$E$ & Date & JD & $T_{\rm{s}}$ (d) & $N_{\rm{NO}}$ & E & Date & JD & $T_{\rm{s}}$ (d)& $N_{\rm{NO}}$\\
\hline
\endhead
\hline
\endfoot
\hline
\hline
\endlastfoot

0 &	1992/04/27 & 2448740 & 49.6	& & 90 & 2003/12/18 & 2452992 & 45.4 & 3-4 \\
3 & 1992/09/05 & 2448871 & 49.0 & & 91 & 2004/01/28 & 2453033 & 44.0 & 4 \\
6 & 1993/01/13 & 2449001 & 44.1 & & 91 & 2004/01/28 & 2453033 & 43.3 & \\ 
7 & 1993/02/27 & 2449046 & 44.5 & & 92 & 2004/03/13 & 2453078 & 46.5 & 4 \\
8 & 1993/04/17 & 2449095 & 45.0 & $>$4 & 92 & 2004/03/13 & 2453078 & 46.8 & \\ 
9 & 1993/05/29 & 2449137 & 43.6 & $>$4 & 93 & 2004/05/01 & 2453127 & 47.5 & 4 \\
10 & 1993/07/12 & 2449181 & 44.4 & & 94 & 2004/06/16 & 2453173 & 46.3 & 3-4 \\
11 & 1993/08/22 & 2449222 & 43.9 & & 94 & 2004/06/16 & 2453173 & 46.6 & \\ 
12 & 1993/10/14 & 2449275 & 44.0 & & 96 & 2004/09/15 & 2453264 & 47.2 & \\ 
13 & 1993/11/18 & 2449310 & 43.6 & & 97 & 2004/11/05 & 2453315 & 47.8 & \\ 
14 & 1994/01/04 & 2449357 & 42.1 & & 97 & 2004/11/06 & 2453315 & 49.2 & \\ 
15 & 1994/02/15 & 2449399 & 43.1 & & 98 & 2004/12/25 & 2453365 & 48.3 & 3-4 \\
16 & 1994/03/29 & 2449441 & 42.6 & $>$4 & 99 & 2005/02/10 & 2453412 & 49.3 & 3-4 \\
17 & 1994/05/11 & 2449484 & 42.5 & $>$4 & 99 & 2005/02/10 & 2453412 & 49.8 & \\ 
18 & 1994/06/24 & 2449528 & 42.4 & 5-6 & 100 & 2005/04/04 & 2453465 & 52.1 & 4 \\
19 & 1994/08/03 & 2449568 & 42.1 & & 101 & 2005/05/22 & 2453513 & 52.2 & 3-4 \\
20 & 1994/09/15 & 2449611 & 42.1 & & 103 & 2005/09/08 & 2453621 & 52.4 & \\ 
21 & 1994/10/27 & 2449653 & 43.5 & & 104 & 2005/10/29 & 2453673 & 53.5 & \\ 
22 & 1994/12/09 & 2449696 & 45.0 & & 104 & 2005/10/30 & 2453673 & 53.4 & \\ 
23 & 1995/01/25 & 2449743 & 46.9 & $>$3 & 105 & 2005/12/23 & 2453728 & 54.9 &\\ 
24 & 1995/03/14 & 2449791 & 47.3 & 5 & 105 & 2005/12/23 & 2453728 & 55.2 & \\ 
25 & 1995/05/02 & 2449840 & 48.71 & 5 & 105 & 2005/12/24 & 2453728 & 56.0 & \\ 
27 & 1995/08/02 & 2449932 & 50.5 & & 106 & 2006/02/17 & 2453784 & 53.3 & \\ 
28 & 1995/09/29 & 2449990 & 54.4 & 5 & 106 & 2006/02/17 & 2453784 & 54.2 & 4 \\
29 & 1995/11/24 & 2450046 & 59.6 & & 107 & 2006/04/08 & 2453834 & 54.7 & 6 \\
30 & 1996/01/31 & 2450114 & 59.4 & & 108 & 2006/06/06 & 2453893 & 55.5 & \\ 
31 & 1996/03/25 & 2450168 & 55.8 & $>$4 & 108 & 2006/06/07 &2453894&56.1&$>$5 \\
32 & 1996/05/22 & 2450226 & 48.9 & & 111 & 2006/11/20 & 2454060 & 55.2 & \\ 
33 & 1996/07/04 & 2450269 & 47.2 & & 111 & 2006/11/21 & 2454060 & 55.0 & \\ 
36 & 1996/11/22 & 2450410 & 47.0 & & 112 & 2007/01/13 & 2454114 & 54.5 & \\ 
37 & 1997/01/06 & 2450455 & 47.6 & 4 & 112 & 2007/01/14 & 2454114 & 55.0 & \\ 
38 & 1997/03/02 & 2450510 & 48.1 & & 113 & 2007/03/10 & 2454169 & 54.9 & \\ 
39 & 1997/04/15 & 2450554 & 47.6 & 4 & 113 & 2007/03/11 & 2454171 & 53.7 & 6 \\
40 & 1997/06/01 & 2450601 & 46.2 & 4 & 114 & 2007/05/03 & 2454224 & 53.6 & 6 \\
42 & 1997/09/04 & 2450696 & 46.0 & & 118 & 2007/12/03 & 2454438 & 52.9 & \\ 
44 & 1997/12/03 & 2450786 & 45.8 & 3-4 & 118 & 2007/12/04 & 2454438 & 51.4 & \\ 
45 & 1998/01/16 & 2450830 & 44.8 & 3-4 & 119 & 2008/01/20 & 2454486 & 45.6 & \\ 
46 & 1998/03/05 & 2450878 & 44.9 & 3-4 & 120 & 2008/03/03 & 2454529 &46.0& 3-4\\
47 & 1998/04/15 & 2450919 & 46.0 & & 120 & 2008/03/04 & 2454530 & 47.0 & \\ 
48 & 1998/06/01 & 2450966 & 49.7 & 5 & 121 & 2008/04/21 & 2454578 & 49.3 & \\ 
49 & 1998/07/21 & 2451016 & 52.2 & & 121 & 2008/04/23 & 2454580 & 49.2 & 3-4 \\
50 & 1998/09/21 & 2451078 & 52.8 & & 122 & 2008/06/10 & 2454628 & 49.7 & \\ 
51 & 1998/11/06 & 2451124 & 50.6 & & 122 & 2008/06/11 & 2454629 & 49.6 & \\ 
52 & 1998/12/28 & 2451176 & 49.2 & & 125 & 2008/11/07 & 2454778 & 50.1 & 3-4 \\
53 & 1999/02/10 & 2451220 & 49.8 & & 125 & 2008/11/08 & 2454778 & 49.8 & \\ 
54 & 1999/04/07 & 2451276 & 51.6 & $>$4 & 126 & 2008/12/29 & 2454830 & 49.6 &\\ 
55 & 1999/05/24 & 2451323 & 52.6 & 5 & 127 & 2009/02/14 & 2454877 & 48.9 & \\ 
57 & 1999/09/11 & 2451433 & 51.0 & & 127 & 2009/02/15 & 2454878 & 48.2 & 4 \\
58 & 1999/10/31 & 2451483 & 51.0 & & 128 & 2009/04/04 & 2454926 & 51.2 & \\ 
59 & 1999/12/13 & 2451526 & 47.6 & 4-5 & 128 & 2009/04/05 & 2454926 & 47.4 & 4\\
60 & 2000/02/06 & 2451581 & 46.0 & & 129 & 2009/05/29 & 2454981 & 47.0 & 5 \\
61 & 2000/03/18 & 2451622 & 46.2 & 5 & 134 & 2010/01/14 & 2455211 & 46.7 & \\ 
62 & 2000/04/30 & 2451665 & 44.7 & 5 & 135 & 2010/03/01 & 2455257 & 45.8 & \\ 
63 & 2000/06/19 & 2451715 & 46.7 & $>$5 & 135 & 2010/03/01 & 2455257 & 46.6 &\\ 
64 & 2000/08/01 & 2451758 & 47.0 & $>$5 & 136 & 2010/04/14 & 2455301&47.7 & 4 \\
65 & 2000/09/21 & 2451809 & 46.5 & & 136 & 2010/04/20 & 2455307 & 46.0 & \\ 
66 & 2000/11/04 & 2451853 & 47.9 & & 137 & 2010/06/05 & 2455353 & 45.6 & 3-4 \\
67 & 2000/12/21 & 2451900 & 47.3 & & 141 & 2010/12/03 & 2455534 & 45.2 & \\ 
68 & 2001/02/11 & 2451952 & 48.4 & $>$5 & 142 & 2011/01/16 & 2455578 & 45.2 &\\ 
69 & 2001/03/27 & 2451996 & 48.4 & & 143 & 2011/03/04 & 2455624 & 45.8 & 4 \\
70 & 2001/05/17 & 2452047 & 48.6 & 6 & 143 & 2011/03/04 & 2455625 & 48.1 & 4 \\
74 & 2001/11/26 & 2452240 & 48.6 & & 143 & 2011/03/05 & 2455626 & 49.5 & 4 \\
75 & 2002/01/17 & 2452292 & 48.0 & 6 & 144 & 2011/04/22 & 2455674 & 49.5 & 4 \\
76 & 2002/03/01 & 2452335 & 45.0 & & 144 & 2011/04/23 & 2455675 & 49.0 & 4 \\
77 & 2002/04/18 & 2452383 & 45.1 & $>$4 & 144 & 2011/04/24 & 2455676 & 48.9 &4\\
81 & 2002/10/10 & 2452558 & 46.1 & & 145 & 2011/06/11 & 2455724 & 48.3 & \\ 
82 & 2002/12/02 & 2452611 & 46.6 & & 145 & 2011/06/11 & 2455724 & 48.9 & \\ 
83 & 2003/01/23 & 2452663 & 50.0 & & 147 & 2011/09/16 & 2455820 & 50.6 & \\ 
84 & 2003/03/10 & 2452709 & 49.2 & $>$5 & 148 & 2011/11/07 & 2455872 & 51.3 &\\ 
85 & 2003/04/29 & 2452759 & 48.3 & 6-7 & 149 & 2012/01/02 & 2455929 & 53.0 & \\ 
85 & 2003/04/30 & 2452760 & 48.2 & & 149 & 2012/01/02 & 2455929 & 53.9 & \\ 
86 & 2003/06/16 & 2452807 & 47.9 & & 150 & 2012/02/20 & 2455978 & 53.4 & \\ 
88 & 2003/09/20 & 2452903 & 45.9 & & 150 & 2012/02/27 & 2455984 & 53.8 & \\ 
88 & 2003/09/21 & 2452903 & 45.1 & & 152 & 2012/06/11 & 2456089 & 54.4 & \\ 
89 & 2003/10/27 & 2452940 & 43.5 & & 152 & 2012/06/12 & 2456090 & 52.8 & \\ 

\hline
\end{longtable}

\section{Discussion}
\label{sec:3}
\subsection{$O-C$ Behavior, Comparison with Other SU UMa DNe}\label{ssec:oc}

\begin{figure*}
\begin{center}
\FigureFile(100mm,100mm){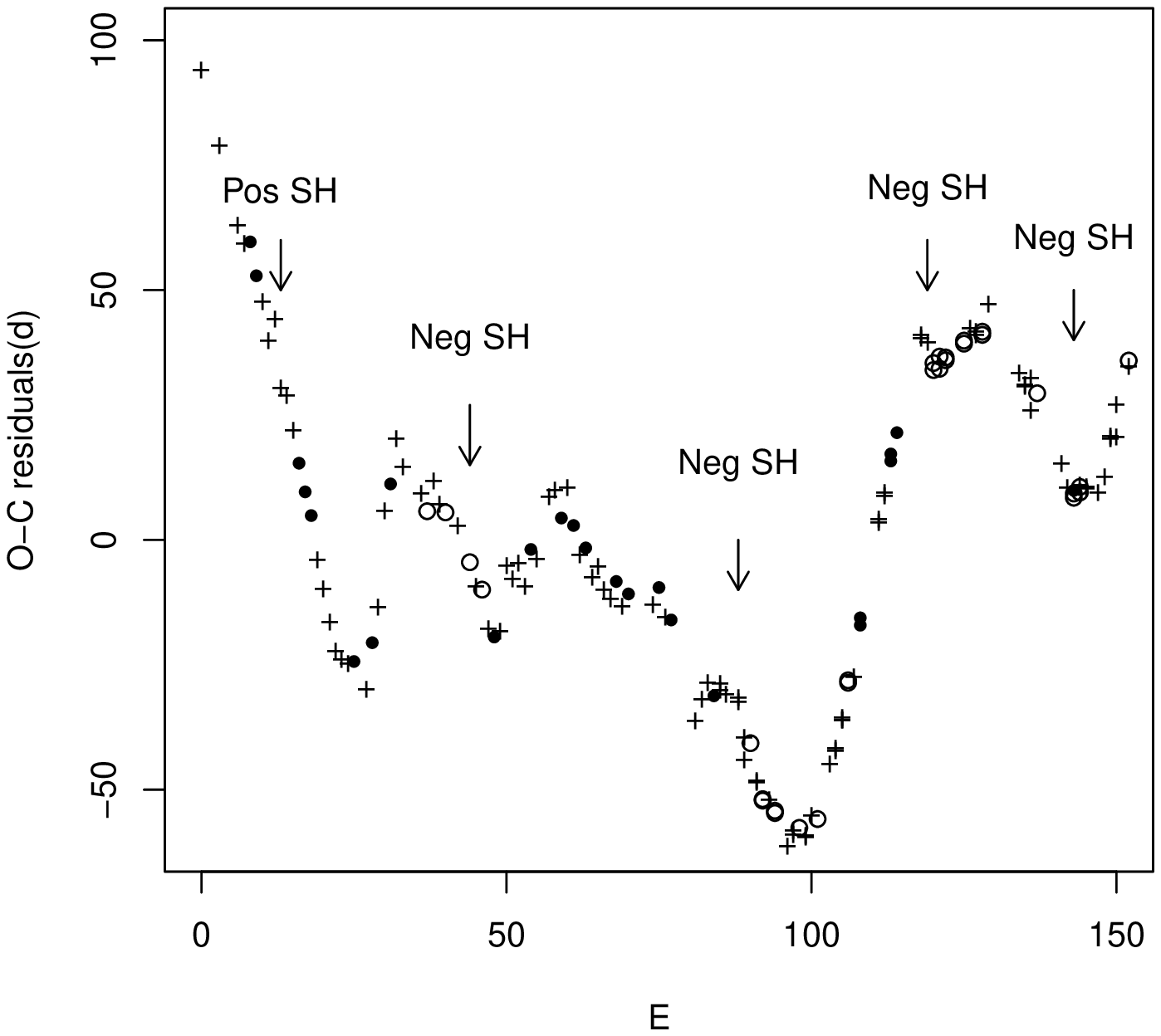}
\end{center}
\caption{$O-C$ diagram for ER UMa with the phases of normal outbursts
and the types of superhumps observed. Open and filled circles correspond to 
L and S normal outburst phases, respectively, supercycles with undetermined 
phases are marked with crosses.}
\label{fig:oc}
\end{figure*}

The $O-C$ diagram for the whole analyzed period is presented in figure 
\ref{fig:oc}. We can see that ER UMa underwent several stages with a relatively
 stable supercycle length (linear portion in the diagram) which
 lasted for 6 -- 30 cycles and terminated by an abrupt period change. The 
local mean supercycle length was determined by the linear fit of each time 
interval with relatively stable supercycle in the $O-C$ diagram. It was shown
 that the supercycle varied from 43.7 at the very beginning of the $O-C$
 diagram to 59.2 d. The local values of the supercycle length are also 
presented in the table \ref{tab:data}. 

\begin{figure*}
\begin{center}
\FigureFile(100mm,100mm){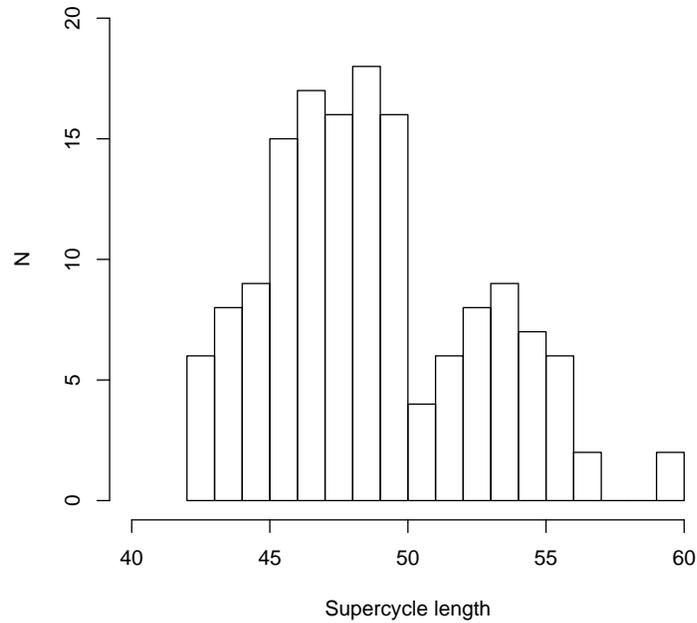}
\end{center}
\caption{Distribution of the supercycle lengths of ER UMa.}
\label{fig:hist}
\end{figure*}

The $O-C$ behavior presented in figure \ref{fig:oc} is reminiscent
 to that of V1159 Ori, shown in figure 2. in \citep{kat01v1159ori}. Although
 one can see such a pattern, for example between $E=60$ and $E=125$ cycles in
 figure \ref{fig:oc}, the time-scale of the variability is significantly longer 
in ER UMa. A detailed analysis of the supercycle variations in SU UMa-type
 stars was presented by (\cite{vog80suumastars}). Although only ordinary SU
 UMa-stars were discussed in \citet{vog80suumastars} (we do not consider 
a comparison with SS Cyg-type DNe since it is out of the scope of this paper), 
 the published $O-C$ diagrams have the same features as ER UMa, particularly of 
SU UMa itself (figure 1 in \cite{vog80suumastars}). Supercycles of SU UMa 
remained stable for relatively long periods of time (10 -- 20 cycles) and
 switched
 randomly between several characteristic values. Nevertheless, while 
\citet{vog80suumastars} has shown that the supercycles in SU UMa possess 3 
characteristic values, in ER UMa there are no preferable supercycle lengths
 as can be seen in figure \ref{fig:hist}.

\subsection{Supercycle Length Change}\label{sec:SC}
\label{sec:SC}

\begin{figure*}
\begin{center}
\FigureFile(100mm,100mm){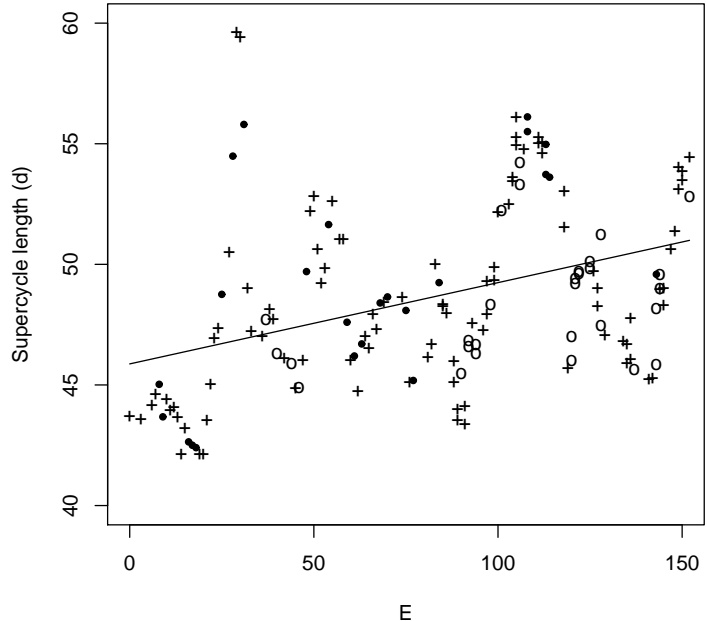}
\end{center}
\caption{Variation of the supercycle length measured by the
 linear fitting of the $O-C$ diagram. The solid line shows
 the secular increasing trend, whose statistical significance was verified by a half-delete jackknife method (\cite{sha95jackknifebook}) with a 90\% confidence. Open and filled circles correspond to 
L and S normal outburst phases, respectively, supercycles with undetermined 
phes marked with crosses.}
\label{fig:sclength}
\end{figure*}

We measured the local lengths of the supercycle by the linear fitting of $O-C$ 
diagram and plotted them against time. Results can be seen in
 figure \ref{fig:sclength} where the supercycle length  change is presented. 
Supercycle lengths ($T_{\rm{s}}$) appeared to change discontinuously and show
 oscillations superimposed on a secular increasing trend. The rate of 
supercycle growth is $dT_{\rm{s}}/dE=0.033(6)~\rm{d}/\rm{cycle}$ and 
 the period derivative is $dT_{\rm{s}}/T_{\rm{s}} = 6.7(6)\times 10^{-4}$,
respectively. We used a half-delete jackknife method described in 
\citep{sha95jackknifebook} to determine reliability of the obtained trend. It 
yielded that the period increase is statistically significant with a 90\% 
confidence. Nevertheless, these formal error estimates should not be
 taken seriously. In \citet{otu12ixdra} it was shown that the period derivative
 for IX Dra is $1.8 \times 10^{-3}$, comparable with that found for ER 
UMa.
 We suppose that it is a typical value of supercycle change rate for ER UMa-type
 stars.  In figure \ref{fig:hist} there is a hint of bimodal distribution
 of the periods of
 supercycle with maxima at 48 and 53 d, respectively.  
\citet{osa95eruma} proposed a model of ER UMa type stars based on the
 TTI theory which successfully explained the light-curves and short
 recurrence times of the superoutbursts assuming extremely high
mass-transfer rate (about 10 times higher than that expected from CV 
evolutionary scenario based on gravitational-wave radiation).
 It also provided a dependence between the period of supercycle and
 mass-transfer rate (presented in figure 1 in \cite{osa95eruma}) and the 
minimum 
possible supercycle length for ER UMa-type stars (about 40 d). According to
 this dependence there are two possible values of $\dot{M}$ for each 
$T_{\rm{s}}$. These two possibilities can be distinguished by considering the
 duty cycle of superoutbursts (ratio of the duration of superoutburst to the 
supercycle length). In our case, the possibility of larger $\dot{M}$ can be 
excluded because the observed duty cycles were much smaller than 0.5. 
Observed variations in the supercycle length of ER UMa correspond to
 the change of mass-transfer rate from $3.8~\dot{M}_{16}$ to $2~\dot{M}_{16}$ 
where $\dot{M}_{16}$ is the mass-transfer rate in units of 
$10^{16}~\rm{g}~\rm{s}^{-1}$
 and the increasing trend in $T_{\rm{s}}$ means that the mass-transfer rate in
 this system is 
decreasing secularly. 
 It is also remarkable that the most significant change of the supercycle and 
 the mass-transfer rate (almost twice) occurred in first break in the $O-C$ 
diagram (JD2449932) right after the initial long stable interval.
 All the further supercycle changes were less dramatic. The period varied
 between 46.2 and 54.1 d which corresponds to a $\sim$ 15 \% change in
 mass-transfer rate. It is also notable that the minimum value of the
 supercycle was not below the shortest achievable supercycle in 
\citep{osa95eruma} and we therefore do not need to assume an additional 
 mechanism as in RZ LMi (\cite{osa95rzlmi}) 

\subsection{Normal Outbursts}\label{ssec:no}

Another important problem in the behavior of ER UMa is the the correlation 
between the length of the supercycle and the number of normal outbursts within
 it. 
For the systems with a high mass-transfer rate (like ER UMa), both the waiting
 time for normal outbursts ($T_{\rm{n}}$) and the waiting time for
 superoutbursts
 ($T_{\rm{s}}$) are approximately proportional to
 $\dot{M}^{-1}$ (\cite{osa95rzlmi}). In this case $T_{\rm{n}}/T_{\rm{s}}$ does 
not depend on 
$\dot{M}$ and the number of normal outbursts between two successive
 superoutbursts should be constant according to the model of ER UMa stars
 proposed by \citet{osa95eruma}. However, as was mentioned in Subsection
\ref{subsec:1no}, ER UMa shows two types of normal outburst phases: a phase 
 with 3-4 normal outbursts observed and with more than 4. In a Type L phase 
presented in the figure \ref{fig:l} ER UMa underwent four normal outbursts
 (marked with arrows) that occurred once in 7 days. In figure \ref{fig:s} 
five normal outbursts can be seen, moreover in the second supercycle the 6-th
 normal outburst near JD2449824 was likely unobserved because of the gap in the
 observations. The time intervals between two subsequent normal outbursts were
 4 days. In figure \ref{fig:oc} and figure \ref{fig:sclength}, filled 
circles correspond to the supercycles with Type S normal outburst phase and open
 circles to the supercycles with Type L phase. If it was impossible to
 determine the type of the phase within the supercycle we mark them with 
crosses. It can be seen from these plots 
that there are no strong correlation between the normal outburst phases and
 the length of the supercycle. Types L and S were observed in any value of the 
supercycle length. Bends on the $O-C$ diagram, or shifts of the supercycle
 period, did not affect the number of normal outbursts. Apparently there 
should be some additional factor, which is responsible for the 
change of the normal outbursts phases. \citet{osa121504cyg}, \citet{ohs12eruma}
  suggested that the type of phase may depend on the 
absence or presence of 
negative superhumps. This possibility is discussed in the following subsection.

\subsection{Negative Superhumps and Outburst Behavior}\label{ssec:negsh}

Negative superhumps are believed to arise from the precession of a tilted 
accretion disk, nevertheless the origin of the disk tilt is still under debate.
 In such cases the hot spot appears not only at
 the disks rim but deeper in the primaries potential, too.
\citet{mon10disktilt} proposed a mechanism based on the hydrodynamic lift. 
Differences between the supersonic velocities of the accretion flow under and
 below the disk cause the appearance of the lift force and make the disk tilted
when the mass transfer is high. The accretion 
stream can thus reach the inner  parts of the accretion disk. Interaction of
 the magnetic fields with rotating plasma was also suggested as a possible
 source of the disk tilt (\cite{mur98ADtilt}, \cite{mur02warpeddisk}, \cite{sma09negativeSH}).  A disk tilt is
 likely to affect the outburst behavior of SU UMa stars, and it is very
 important to study the correlation between the the presence of negative 
superhumps (as an indicator of the tilt of the disks) and outbursting activity.
 In 2011 it became evident that persistent
 negative superhumps in ER UMa were observed even during the 
superoutburst (\cite{ohs12eruma}) while a retrospective search for the
 literature and the available data suggests that they were absent in the past.
 \citet{kat95eruma} in their 
discovery paper reported the detection of positive superhumps in ER UMa. 
No negative superhumps which are considered as characteristic feature of ER
 UMa in the present state, were observed by them in 
1994. The first possible detection of negative superhumps in ER UMa was 
claimed by 
\citet{gao99erumaSH}, but the amplitudes were smaller than in \citet{ohs12eruma}
 and it is likely that, even if the negative superhumps did exist in 1998, they
 were not so strong in 1998 as in 2011. Since then the presence
 of the superhumps with period less than the orbital one was confirmed by
 \citet{zhao06er}, \citet{kju10eruma}. In our retrospective study we also 
confirmed the presence of negative superhumps in 2010 in the AAVSO data. It 
appears likely that the negative superhumps were present since 2006. 

\subsection{Correlation with Normal Outburst Phase}

TTI theory suggests that a mass-transfer to a tilted disk is expected to reduce
 the number of normal outbursts (\cite{osa121504cyg}). Since the
 accretion stream in such cases can reach the inner parts of the disk, material
 is accumulating not primarily on the disks rim. It leads to the reduction of 
``outside-in'' outbursts.  
 Here we examine the correlation between the normal outburst phase and the
 presence of negative superhumps. In the figure \ref{fig:oc} we mark the types
 of superhumps observed by the authors mentioned above. One can see that the
 presence of negative superhumps seems to depend strongly on the phase of the
 normal outbursts. The first possible detection of negative superhumps was near
 the $E=44$ on figure \ref{fig:oc} when the first L stage was observed. The next
 one, claimed by \citet{zhao06er} was on 2004 February 29 which corresponds to
 the $E=88$, just at the transition from S to L phases. 
 Negative superhumps observed by \citet{kju10eruma} in 2010 were in $E=119$
 at the diagram, also at the beginning of L phase. 
According to the detailed light curves of ER UMa provided by the observational
 campaign of the VSNET collaboration \citep{VSNET} negative superhumps have been
 observed since the 2011 ($E=135$) and four normal outbursts occurred within 
each supercycle.
 It confirms that the phase of the normal outbursts can be strongly affected 
by the presence of negative superhumps, and negative superhumps can indeed 
reduce the number of normal outbursts. So we can now assume that Type L phase
 can be regarded as ``negative superhump" phase.

\subsection{Correlation with the Supercycle length}
 According to figure \ref{fig:oc}, positive and negative superhumps were 
observed both on the descending and ascending branches of the $O-C$ diagram, and
 they do not correlate with the supercycle length. Negative superhumps
were with no doubt observed since 2011. During this period $O-C$ underwent 
a bend near $E=150$ but it had no affect on the presence of negative 
superhumps. 
In the TTI model, the next superoutburst occurs when the disk stores a certain 
amount of angular momentum ($J_{crit}$), and supercycle is not expected to 
change if $\dot{J}_{crit}$ from the secondary is constant regardless of the
 point of accretion on the disk. A disk tilt therefore is not expected to affect
 the supercycle length and our analysis confirmed this expectation.

\section{Conclusions} 
In this paper we examined data of about 20 years of observations of ER
 UMa. We followed the evolution of the change in supercycle length, the 
behavior of normal outbursts and the types of superhumps observed. We found the 
oscillations of the supercycle length superimposed with a secular increasing
 trend and associated it with a gradual change of the mass-transfer rate in
 the binary system. The estimated period derivative was 
$dT_{\rm{s}}/T_{\rm{s}} = 6.7(6)\times 10^{-4}$. We also found that ER UMa 
shows two types of normal
 outburst phases with 3-4 and with 5 or more normal outbursts within the
 supercycle. We also examined the possible correlation between the supercycle
 length, normal outburst phase and the presence of negative superhumps and 
found a strong dependence between the last two. A reduced number of
 normal outbursts was observed when negative superhumps were detected. It 
confirms the assumption of \citet{osa121504cyg} that negative
 superhumps tend to suppress the normal outbursts. No correlation between the 
length of the  supercycle and other characteristics was found.

\vskip 3mm

This work was supported by the Bilateral International Exchange Program (2012) 
offered by the Kyoto University and Global COE ``The Next Generation of Physics,
 Spun from Universality and Emergence'' from the Ministry of Education,
 Culture, Sports, Science and Technology (MEXT) of Japan. 

Dr. Shugarov acknowledge the Slovak Academy of Sciences grant VEGA No.
 2/0038/10, grant RFBR No. 09-02-00225 and NSh-2374.2012.2 (Russia).

We acknowledge with thanks the variable star observers from the American
 Association of Variable Stars Observers, Variable Stars NETwork, Association 
Fran\c{c}aise des Observateurs d'Etoiles Variables, Northern Sky Variability 
Survey and Variable Star Observers League in Japan.
We also acknowledge the group of Elena Pavlenko for the data presented and 
would like to thank Yoji Osaki for valuable comments.

\end{document}